\documentclass[10pt,a4paper]{article}
\usepackage[letterpaper,text={6.5in,9in},centering]{geometry}
\usepackage{graphicx}
\usepackage{bm}
\usepackage{verbatim}
\usepackage{color}
\usepackage{amsmath}
\usepackage{float}
\usepackage{amsfonts}
\usepackage[makeroom]{cancel}
\usepackage[autostyle]{csquotes}
\usepackage{graphicx}
\usepackage{ulem}
\usepackage[toc,page]{appendix}
\usepackage{cite} 
\graphicspath{ {images/} }
\newcommand{\tr}{\textrm{tr}}

\title{Properties of Infinite Nuclear Medium from QCD Sum Rules and the Neutron Star-Black Hole Mass Gap}
\bigskip
\author{Bijit Singha$^{a}$\footnote{email : bijit.singha89@gmail.com} , Debasish Das$^{b}$\footnote{email : debasish.das@saha.ac.in,dev.deba@gmail.com}, Leonard S. Kisslinger$^{c}$\footnote{email : kissling@andrew.cmu.edu}\\
\medskip
$^a$Nomura, Hiranandani Gardens, Powai, Mumbai, MH 400076, India. \\
$^b$Saha Institute of Nuclear Physics, A CI of Homi Bhabha National Institute,\\
 1/AF, Bidhan Nagar, Kolkata, WB 700064, India.\\
$^c$Department of Physics, Carnegie Mellon University, Pittsburgh, PA 15213, USA.\\
}
\date{}
\begin{document}
\maketitle

\begin{abstract}
A non-perturbative framework is provided to connect QCD with nuclear phenomenology in the intermediate density regime. Using QCD Sum Rules, in-medium scalar and vector self-energies of nucleons are calculated as functions of the density of an infinite nuclear medium. The self-energies are used in the relativistic mean field theory lagrangian of a high-density nuclear medium to find the binding energy of in-medium nucleons and the value of light quark condensate, $\langle \bar{q} q \rangle_{\rm{vac}} = -~(0.288 ~\rm{GeV})^3$, in the Borel-improved resummation scheme. The critical mass of an ideal neutron star is obtained by coupling a uniform saturation energy density of cold, dense nuclear matter to Einstein equation in hydrostatic equilibrium. Since it is less likely for a neutron star core to avoid deconfinement and enter the rigid vector repulsion phase where the speed of sound can smoothly approach from conformal to causal limit, a gap should exist in the stellar mass spectrum, $[3.48M_\odot, 5.47M_\odot]$,
\noindent where it would be rare to find any isolated, cold, non-rotating neutron star or a black hole.
\end{abstract}

\vspace{1mm}
\noindent
Keywords: Nuclei, Sum Rules, Nuclear Matter, Neutron Star, Black Hole, Mass Gap, General Relativity and Quantum Cosmology; Nuclear Astrophysics; Astrophysics - High Energy Astrophysical Phenomena

\section{Introduction}
The fundamental degrees of freedom in Quantum Chromodynamics(QCD) are quarks~\cite{Riordan:1992hr} and gluons~\cite{Ellis:2014rma} which have never be observed in isolation in any physical experiment. The physical states and the interactions that we prevalently observe in experiments are QCD bound states and the residual part of QCD interactions, namely hadronic interactions~\cite{Marciano:1977su, Shuryak:1980tp, Kisslinger:2014uda, Adams:2005dq, Adcox:2004mh, CBMFAIRQCD, Citron:2018lsq}. While the fluctuations of QCD vacuum~\cite{EIC2022} are responsible for making the nucleons hugely massive, the interaction between these nucleons occurs through the exchange of mesons which are rather QCD bound states. In fact, a renormalizable field theory can be consistently formulated in terms of Lorentz-invariant lagrangian density written in terms of hadrons as fundamental degrees of freedom (d.o.f.)~\cite{Walecka1974, Walecka1997} to provide a number of successful predictions in nuclear physics without even considering the color interactions as laid out in QCD.\\

A quantitative understanding of this quark-hadron duality as well as establishing its accuracy over all energy scale still remains a challenging issue. For sufficiently high energy, this duality sets in with asymptotic freedom~\cite{Coleman1973, Weinberg1973, Gross1973, Gross1974} which makes the QCD calculations trivial. But non-perturbative effects start to dominate at an energy around the nuclear mass. Although the pion-pion and pion-nucleon interactions are weak for small momenta at that scale due to spontaneous breaking of chiral symmetry~\cite{PiPi,PiN}, the nucleon-nucleon interactions become non-perturbative. The issue becomes more complex when, at the same energy scale, we try to understand this duality in the context of the thermodynamic properties of dense nuclear matter. This specific problem behooves to be solved only analytically, since any deviation from or violation of quark-hadron duality is essentially a Minkowskian phenomenon and numerical Euclidean approach such as Lattice QCD suffers from a sign problem~\cite{SignProblem1,SignProblem2} that typically occurs for a non-zero chemical potential in the presence of a fermionic background such as a dense nuclear medium. Moreover, the characteristic energy scale of QCD and that of nuclear physics are different by orders of magnitude. Lightest hadrons can be as massive as hundreds of MeV while the typical binding energy that dictates the nuclear phenomena is only around a few MeV. Also, the physics of exotic states of matter in high temperature or large compression, where the QCD d.o.f. start to predominate hadronic d.o.f., is not fully developed yet. All these impose serious challenges in utilizing quark-hadron duality to explore the connections between the QCD theory and nuclear physics observables as well as to study astrophysical objects like neutron star~\cite{Baym:1978jf,McLerran:2018hbz,Kisslinger:2021ach}.\\

A popular approach to go around these challenges is to frame an effective lagrangian of strong interaction by exploiting the symmetry properties of QCD in the chiral limit~\cite{Weinberg1967, Weinberg1968, Coleman1969First, Coleman1969Second, Dashen1969First, Dashen1969Second}, although utility of this approach to predict the behavior of dense nuclear matter is still very limited. Attempts have also been made to map the behavior of strongly coupled QCD medium to a higher dimensional theory of gravity using AdS-CFT duality~\cite{Teramond2005, Erlich2005, Rold2005, Karch2006, Brodsky2015, Alho2013, Jarvinen2022}. But this approach faces challenges too because, unlike CFT, (a) QCD is not superconformal, (b) QCD has confinement, (c) the number of QCD colors as well as the 't Hooft QCD coupling are not infinite.\\

In this work, we exploit quark-hadron duality to successfully predict the physical properties of saturated nuclear matter and of an ideal neutron star that is comprised of such matter. The methodology we use is called QCD Sum Rules~\cite{Shifman1979First, Shifman1979Second, ReindersQuarkProp, LSKBS1, LSKBS2}. QCD Sum Rules for nucleons in nuclear medium have been investigated years ago~\cite{Cohen1991, Cohen1992, Cohen1993, Cohen1994, Cohen1995} with a number of important predictions such as positive vector self-energy, the reduction of effective nucleon mass arising from the reduction of the in-medium light-quark condensate its vacuum value \textit{etc.} which matches with the conclusions of our work. These articles observed strong dependence of the effective nucleon mass on the Borel mass which is a redundant parameter from the phenomenological point of view. In our  work, the Borel mass was fixed using the condition that the effective nucleon mass will approach the physical nucleon mass in the absence of the surrounding nuclear medium. Moreover, we derive the value of the light quark condensate in vacuum, effective mass of a nucleon as a function of medium density, and the saturation curve for nuclear matter. All these results show similar quantitative features as found in the Quantum Hadrodynamics(QHD) calculations by Walecka \textit{et. al.} in~\cite{Walecka1974, Walecka1997}.\\

The importance of this work lies in the fact that it provides an intuitive, systematic, and non-perturbative framework to connect QCD to nuclear phenomenology, especially in intermediate density regime. This work provides an effective, non-perturbative way to match the phenomenology arising from QCD lagrangian with the predictions of the long-range, strongly coupled effective field theory of QHD. Furthermore, this framework has the capacity to provide equation of state for nuclear matter with all densities higher than observed terrestrial density, -- essential for the estimation of the physical properties of neutron stars from first principles QCD calculations.\\

The structure of this paper is as follows. In Sec.~\ref{Sec:2ptCorr}, we start explicitly with the light-quark d.o.f. of QCD to write an Operator Product Expansion(OPE) of the in-medium nucleon two-point function. In Sec.~\ref{Subsec:QuasiN}, we propose a phenomenological model of in-medium nucleons consistent with hadron scattering observations. The Borel-transformed OPE and the model is then compared in Sec.~\ref{Subsec:BT} to give us the expressions for scalar and vector self-energies of the nucleon in terms of quark condensates in Sec.~\ref{SelfE}. The self-energies are used in the mean-field QHD lagrangian density of Sec.~\ref{SaturationCurve} to obtain the saturation curve, value of the coupling parameters and the value of vacuum light-quark condensate in Sec.~{\ref{Results}. Sec.~\ref{Ch_NS} uses these information in Tolman-Oppenheimer-Volkoff equations to obtain the maximum masses of a neutron star for the scenarios when the speed of sound in neutron star core approaches the conformal and causal limit. Sec.~\ref{Ch_Regime} discusses the possibility of a universal gap in the stellar mass spectrum where it would be rare to find any isolated, cold, non-rotating neutron star or a black hole.

\section{Two-point Correlator of Nucleons in Nuclear Medium}
\label{Sec:2ptCorr}
In QCD Sum Rules, we attack the nuclear bound state problem from the short distance side and gradually move to larger distances where asymptotic freedom starts to break down and exploit the non-trivial structure of the QCD vacuum signalled by the emergence of power corrections. Quantitatively, we represent the hadrons propagating in nuclear matter in terms of their interpolating quark current at large virtualities. Then we construct a two-point correlator using the hadron operators. We treat the correlator in the OPE framework where long and short distance contributions are dealt differently: the former is represented as Wilson's coefficients and are evaluated using perturbative QCD, while the latter entails infrared behavior of the Green's functions of quarks and gluons and is represented by various condensates. In our calculation, we consider only the identity operator and the light-quark condensate, which are the operators for the OPE up to mass dimension three.\\

In order to write QCD Sum Rules for nucleons in nuclear medium, we start with a color-singlet hadron current that couples maximally to a nucleon
\begin{eqnarray}
\label{DefHadronCurrent}
\eta_N(x) &=& \epsilon_{abc} \left[  
q^{aT}(x) C \gamma_\mu q^b(x)
\right] \gamma_5 \gamma^\mu q^c(x),
\end{eqnarray}
where $C$ denotes the charge conjugation matrix, $T$ denotes transpose in Dirac space, $q$ denotes light quark field with $SU(2)$ isospin symmetry, $a, b, c$ denote color indices and $\epsilon_{abc}$ is the totally antisymmetric tensor on the index subset of the three color indices. The choice of current for a given $J^{PC}$ is not unique but it is chosen in such a way that the coupling to nucleon intermediate state is maximized while the contribution of the higher order states to the correlation function is negligible. Sensitivity to the choice of current in Eq.~(\ref{DefHadronCurrent}) is discussed in~\cite{Cohen1994}.\\

We consider the two-point correlator in momentum space comprised of the time-order product of the local hadron current and its Hermitian conjugate
\begin{eqnarray}
\label{DefTwoPointCorr}
\Pi_2^N (p) &=& i \int d^4 x~ e^{ip.x}~ \langle 0 \vert T \big[
\eta_N (x) \bar{\eta}_N (0)
\big]
\vert 0 \rangle~,
\end{eqnarray}
where the expectation value is taken over physical, non-perturbative vacuum state. We use Eq.~(\ref{DefHadronCurrent}) in Eq.~(\ref{DefTwoPointCorr}) and using the light-quark propagator in fixed point gauge to the first order in the light quark mass in spacetime coordinate in the presence of background quark field\cite{ReindersQuarkProp}
\begin{eqnarray}
\label{QPropagatorSRMain}
\left[ S^q_{ab}(x) \right]_{\alpha \beta} = \frac{i}{2 \pi^2} \delta_{ab} \frac{\cancel{x}_{\alpha \beta}}{(x^2)^2}  -  \delta_{ab} \delta_{\alpha \beta} \frac{m}{4 \pi^2 x^2} - \frac{\delta_{ab}}{12} \left[
\langle \bar{q} q \rangle_{\rho_N} \delta_{\alpha \beta} + \langle \bar{q} \gamma_\lambda q \rangle_{\rho_N} \gamma^\lambda_{\alpha \beta}
\right] + \cdots ,
\end{eqnarray}
to write the OPE:
\begin{eqnarray}
\label{Pi2NMain}
\Pi_2^N (p) &=& 12 i \int d^4 x~ e^{ip.x}~ \tr\Big[
\gamma_\mu \Big\{ 
\frac{i}{2 \pi^2} \frac{\cancel{x}}{(x^2)^2} - \frac{m}{4 \pi^2 x^2} - \frac{1}{12}\big[
\langle q^\dagger q \rangle_{\rho_N}\cancel{u} + \langle \bar{q} q \rangle_{\rho_N}
\big]
\Big\} \gamma_\nu \nonumber\\ &&
 \qquad \Big\{ 
\frac{i}{2 \pi^2} \frac{\cancel{x}}{(x^2)^2} + \frac{m}{4 \pi^2 x^2} + \frac{1}{12}\big[
\langle q^\dagger q \rangle_{\rho_N}\cancel{u} + \langle \bar{q} q \rangle_{\rho_N}
\big]
 \Big\}
\Big]\nonumber\\&& 
\qquad \qquad \times
\left[ \gamma^\mu \Big\{
\frac{i}{2 \pi^2} \frac{\cancel{x}}{(x^2)^2} + \frac{m}{4 \pi^2 x^2} + \frac{1}{12}\big[
- \langle q^\dagger q \rangle_{\rho_N}\cancel{u} + \langle \bar{q} q \rangle_{\rho_N}
\big]
\Big\} \gamma^\nu  \right] ~.
\end{eqnarray}
In Eq.~(\ref{Pi2NMain}), $u^\mu$ denotes the four velocity of the nuclear medium and $u^\mu = (1,\textbf{0})$ in the rest frame of the nuclear medium. This additional four vector is introduced in the formalism keeping in mind that there exists a preferred reference frame for the nuclear matter which is its own rest frame and observations in all other reference frames are connected through its Lorentz transformations. It is also to be noted that, in writing the light-quark propagator in Eq.~(\ref{QPropagatorSRMain}), we have ignored the gluonic contributions which holds true for an expansion of the propagator for sufficiently short distance.\\

Eq.~(\ref{Pi2NMain}) can further be simplified to give us the OPE to leading order:
\begin{eqnarray}
\label{Pi2NFinal}
\Pi_2^N (p) = \Pi_s \left(p^2, p.u \right) + \Pi_q\left(p^2, p.u \right) \cancel{p} + \Pi_u \left( p^2, p.u \right) \cancel{u} 
\end{eqnarray}
with
\begin{eqnarray}
\label{OPE1}
\Pi_s(p^2, p_0) &=& - \frac{1}{4 \pi^2}  P^2 \ln P^2 \langle \bar{q} q \rangle_{\rho_N} + \cdots ~, \\
\label{OPE2}
\Pi_q(p^2, p_0) &=& - \frac{1}{64 \pi^4} (P^2)^2 \ln P^2 + \frac{1}{3 \pi^2} p_0 \ln P^2 \langle q^\dagger q \rangle_{\rho_N} + \cdots~, \\
\label{OPE3}
\Pi_u(p^2, p_0) &=& -\frac{2}{3 \pi^2} P^2 \ln P^2  \langle q^\dagger q \rangle_{\rho_N} + \cdots~,
\end{eqnarray}
after omitting all the power-divergent terms which vanish anyway after Borel transformation of the two-point function. The above expressions match the OPE derived in~\cite{Cohen1991} with light-quark isospin symmetry assumed in our work. We would like to mention here that the two-point correlator $\Pi_2^N(p)$ can be expanded into a much general form in Dirac space. But the symmetries of Lorentz covariance, time reversal and parity dictate that the two-point correlator comprises three distinct structures as shown in Eq.~(\ref{Pi2NFinal}) and this is true to all orders of OPE in Eq.~(\ref{Pi2NMain}). Moreover, in the limit where the density of nuclear medium is zero, $\Pi_u(p^2,p_0) \rightarrow 0$  and $\Pi_q$ become a function of $P^2$ only. This overlaps with the fact that the symmetry of the two-point correlator in vacuum contains only two structures: scalar and $\cancel{p}$. 
\section{Phenomenological Side of the Sum Rule}
\label{Sec:pheno}
\subsection{Quasinucleons in Large Nuclear Medium}
\label{Subsec:QuasiN}
In the previous section, we derived the expression for Fourier transform of the nucleon two-point correlation function using OPE. Lehmann-Kallen spectral representation tells us that the analytic structure of this correlation function in complex-$p^2$ plane should have an isolated simple pole at the mass of a quasi-nucleon state. In view of this fact and with the help of empirical data as well as theoretical features of the hadron scattering phenomena, we now try to come up with a phenomenological model of the two-point correlation function at an intermediate energy. In line of Dirac-Brueckner-Hartee-Fock (DBHF) approach, we assume weak three-momentum dependence of the in-medium self-energies for the bound states as well as low-lying continuum states~\cite{Cohen1995,Walecka2004-1}. \\

We assume that the two-point function will have a pole at the physical nucleon mass to write\cite{Cohen1995}
\begin{eqnarray}
\label{Pheno2P}
\Pi_N(p) &=&- ~ \frac{\lambda_N^{*2}}{(p^\mu - \Sigma^\mu_V) \gamma_\mu - \left( M_N + \Sigma_S  \right)} + {\rm{continuum}}~.
\end{eqnarray}
where $\lambda_N^*$ is the coupling of the nucleon current to the physical quasinucleon in the nuclear medium, $\Sigma^\mu_V$ and $\Sigma_S$ are the in-medium vector and scalar self-energies. From the model above, we expect to retrieve a few essential features such as attractive scalar and repulsive vector potential of several hundreds of MeV cancelling each other in such a way that the binding energy of a nucleon turns out to be only a few MeV eventually. It is important to note here that the scalar and vector potentials are physically observed only in combinations and not individually. Additionally, cancellation of the imaginary part of scalar and vector potential is expected in order to achieve a stable quasinucleon state. \\

We can further expand the vector self-energy to write
\begin{eqnarray}
\Sigma^\mu_{V} = \Sigma_V u^\mu + \Sigma'_{V} q^\mu ~.
\end{eqnarray}
We neglect $\Sigma'_V$ because of weak $q$-dependence of this term and the negligible contributions from higher mass excitations. Doing this for a nuclear medium in the rest frame, we square the denominator of Eq.~(\ref{Pheno2P}):
\begin{eqnarray}
\label{pSqMinusMuSq}
(p^\mu - \Sigma^\mu_V)^2 - M_N^{*2} = p^2 - \mu^2,
\end{eqnarray}
where
\begin{eqnarray}
\label{muSq}
\mu^2 =  M_N^{*2} + 2 p_0 \Sigma_V - \Sigma_V^2~.
\end{eqnarray}
Finally, in the phenomenological side, we use Eq.~(\ref{Pheno1}, \ref{pSqMinusMuSq}, \ref{muSq}) together to derive the following expressions for different tensor structures as in Eq.~(\ref{Pi2NFinal}):
\begin{eqnarray}
\label{Pheno1}
\Pi_s(p^2, p_0) &=& - \lambda_N^{*2} \frac{M_N^*}{p^2 - \mu^2},\\
\label{Pheno2}
\Pi_q(p^2, p_0) &=& - \lambda_N^{*2} \frac{1}{p^2 - \mu^2}, \\
\label{Pheno3}
\Pi_u(p^2, p_0) &=& - \lambda_N^{*2} \frac{\Sigma_V}{p^2 - \mu^2}~.
\end{eqnarray}

\subsection{Borel Transformation}
\label{Subsec:BT}
Because of the infrared slavery in QCD, insertion of more bubbles in the QCD Feynman diagrams leads to softer momenta where the coupling constant starts to become increasingly large. Thus, the IR region of the loop integral becomes more and more important and running coupling assumes leading logarithms in its expression. These different powers of logarithms gives us factorial divergence in the expansion of the correlation function. In these cases, we try to achieve rapid convergence for a resummed Borel series of the correlator. We make a Borel transformation of the form shown here, where we consider sufficiently high moment of the correlator and a high momentum, where only the contribution from the lowest resonance predominates all other resonances in the channel. Additionally, spurious power divergences could appear in OPE but they vanish with Borel transform. Hence, all such terms were already ignored in the OPE of time-ordered product.\\

Borel transform of a function $f(P^2)$ is defined as:
\begin{eqnarray}
\label{BorelT}
\mathcal{B}_{M^2} \left[
f(P^2)
\right] =
\lim_{P^2, n \rightarrow \infty, P^2/n = M^2}
\frac{\left(P^2\right)^{n+1}}{n!} \left(
\frac{-d}{dP^2}
\right)^n f(P^2)~.
\end{eqnarray}
Using this definition, we derive the following formulae:
\begin{eqnarray}
\mathcal{B}_{M^2} \left[ \frac{1}{p^2 - \mu^2} \right] &=& - e^{-\mu^2/M^2},\\
\mathcal{B}_{M^2} \left[ \ln P^2 \right] &=& -M^2, \\
\mathcal{B}_{M^2} \left[ P^2 \ln P^2 \right] &=& M^4, \\
\mathcal{B}_{M^2} \left[  (P^2)^2 \ln P^2 \right] &=& - 2 M^6~,
\end{eqnarray}
which we use in Eq.~(\ref{OPE1}-\ref{OPE3}) and Eq.~(\ref{Pheno1}-\ref{Pheno3}) to write the OPE tensor structures in terms of their phenomenological counterparts
\begin{eqnarray}
\label{PiSBorelTransformed}
\lambda_N^{*2} M_N^* e^{-\mu^2/M^2} &=& - ~\frac{\langle \bar{q} q \rangle_{\rho_N}}{4 \pi^2}~ M^4,\\
\label{PiqBorelTransformed}
\lambda_N^{*2} e^{-\mu^2/M^2} &=& \frac{M^6}{32 \pi^4} - \frac{p_0}{3 \pi^2} \langle q^\dagger q \rangle_{\rho_N} M^2,\\
\label{PiuBorelTransformed}
\lambda_N^{*2} \Sigma_V e^{- \mu^2/M^2} &=& \frac{2}{3 \pi^2}  \langle q^\dagger q \rangle_{\rho_N} M^4~.
\end{eqnarray}

\subsection{Self-energies in Terms of Quark Condensate}
\label{SelfE}
Eq.~(\ref{PiSBorelTransformed}, \ref{PiqBorelTransformed}, \ref{PiuBorelTransformed}) have quark condensates in their expressions which arise from the fact that the chiral symmetry of QCD gets spontaneously broken in QCD vacuum with the condensates emerging as the order parameters of this phenomenon. Restoration of chiral symmetry may happen, for example, in the core of a neutron star where the hadron medium has sufficiently high density. In scenarios where the environment is not this much extreme, the chiral symmetry can be partially restored in the presence of any hadron medium. This leads to a change in the value of quark condensate, $\langle \bar{q} q \rangle_{\rho_N}$, in presence of a nuclear medium of density $\rho_N$  and is given by~\cite{Drukarev1990, Drukarev1994},
\begin{eqnarray}
\label{condinmedium}
\langle \bar{q}q \rangle_{\rho_N} =
\langle \bar{q}q \rangle_{\rm{vac}} \left[
1 - \frac{\rho_N \sigma_N}{f_\pi^2 m_\pi^2}
\right],
\end{eqnarray} 
to leading order. Here $f_\pi$ is the pion decay constant and $\sigma_N$ is the pion-nucleon sigma term~\cite{Reya1974}. It can be shown that $f_{\pi}^2 m_{\pi}^2$ is related to the symmetry breaking term of the QCD lagrangian through Gell-Mann-Oakes-Renner relation:
\begin{eqnarray}
\label{piondecayconsttocond}
f_{\pi}^2 m_{\pi}^2 \approx - 2 m_q \langle \bar{q} q \rangle_{\rm{vac}}~,
\end{eqnarray}
where $m_q$ is the average current mass of the up and down quarks. Eq.~(\ref{condinmedium}) and (\ref{piondecayconsttocond}) together gives us
\begin{eqnarray}
\label{condinmediumfinal}
\langle \bar{q}q \rangle_{\rho_N} =
\langle \bar{q}q \rangle_{\rm{vac}}
+ \frac{\rho_N \sigma_N}{2 m_q}~.
\end{eqnarray}
Furthermore, $\langle q^\dagger q\rangle_{\rho_N}$ is related to the net nucleon density~\cite{Jin1994,Jin1995}:
\begin{eqnarray}
\label{qDaggerq}
\langle q^\dagger q\rangle_{\rho_N} = \frac{3}{2} \rho_N~.
\end{eqnarray}
Using Eq.~(\ref{qDaggerq}) in Eq.~(\ref{PiqBorelTransformed}), we get
\begin{eqnarray}
\lambda_N^{* 2} e^{-\mu^2/M^2} = \frac{M^6}{32 \pi^4} - \frac{\left(k_F^2 + M^{*2}_N\right)^{1/2} + \Sigma_V}{3 \pi^2} \times \left( \frac{3}{2} \rho_N M^2 \right)~.
\end{eqnarray}
Now we use Eq.~(\ref{PiSBorelTransformed}) in the RHS of the above expression to get
\begin{eqnarray}
- \frac{\langle \bar{q}q \rangle_{\rho_N}}{4 \pi^2} \frac{M^4}{M_N^*} = \frac{M^6}{32 \pi^4} - 
\frac{\left[ \left(\frac{k_F}{M_N^*}\right)^2 + 1\right]^{1/2} + \frac{\Sigma_V}{M_N^*}}{3 \pi^2} \times \left( \frac{3}{2}\rho_N M_N^* M^2 \right)~.
\end{eqnarray}
We can express the baryon density($\rho_N$) in terms of Fermi momentum ($k_F$):
\begin{eqnarray}
\label{Density}
\rho_N = \frac{\gamma}{(2\pi)^3} \int_0^{k_F} d^3 \vec{k} = \frac{\gamma k_F^3}{6 \pi^2}~.
\end{eqnarray}
where $\gamma$ is the degeneracy factor of the nucleon and it is 4 for nucleon with unbroken isospin symmetry ($2$ for spin and $2$ for isospin). Writing $(k_F/M_N)$ as $x$ and $(M_N^*/M_N)$ as $y$, we get
\begin{eqnarray}
\label{EoS}
- \frac{1}{2} \left( \frac{M}{M_N}  \right)^2 \frac{1}{y} \left[
 \frac{\langle \bar{q} q\rangle_{\rm{vac}}}{M_N^3} + \frac{2}{3 \pi^2} \left( \frac{\sigma_N}{2 m_q} \right) x^3  
\right] = \frac{1}{16 \pi^2} \left(\frac{M}{M_N}\right)^4 - \frac{2x^3y}{3 \pi^2} \left[\frac{\Sigma_V}{M_N^*} + \left( 1 + \frac{x^2}{y^2} \right)^{1/2} \right]~.
\end{eqnarray}
We get the ratio $\Sigma_V/M_N^*$ by dividing Eq.~(\ref{PiuBorelTransformed}) by Eq.~(\ref{PiSBorelTransformed}),
\begin{eqnarray}
\label{RatioSigmaV}
\frac{\Sigma_V}{M_N^*} &=& -~ \frac{4}{\left( \frac{\langle \bar{q} q \rangle_{\rm{vac}}}{\rho_N} + \frac{\sigma_N}{2 m_q} \right)}~.
\end{eqnarray}
We define 
\begin{eqnarray}
\label{def_r}
r = - ~ \frac{\langle \bar{q} q \rangle_{\rm{vac}}}{M_N^3}.
\end{eqnarray}
From~\cite{Gasser1991, Cohen1995}, we assume $\sigma_N = 45$ MeV and $m_q = 2.5$ MeV to write
\begin{eqnarray}
\label{EffVectorPot}
\frac{\Sigma_V}{M_N^*} = -~\frac{4}{9 - 3 \pi^2 r/2x^3}~.
\end{eqnarray}
Now, let's use the condition that in the limit of zero density, the effective nucleon mass equals the physical nucleon mass: $y \rightarrow 1$ as $x \rightarrow 0$. This gives us the Borel mass parameter:

\begin{eqnarray}
\label{ratioBorel}
\frac{M}{M_N} = \left[- \frac{8 \pi^2}{M_N^3} \langle \bar{q} q \rangle_{\rm{vac}} \right]^{1/2}  = [8 \pi^2 r]^{1/2}.
\end{eqnarray}
Using Eq.~(\ref{def_r}), Eq.~(\ref{EffVectorPot}) and Eq.~(\ref{ratioBorel}) in Eq.~(\ref{EoS}) and defining
\begin{eqnarray}
z = r - \frac{6}{\pi^2}x^3,
\end{eqnarray}
we can finally write
\begin{eqnarray}
\label{FinalEoSEq}
z -ry + \frac{x^3 y}{6 \pi^4 r} \Big[
\big( x^2 + y^2 \big)^{1/2} + \frac{8x^3 y}{3 \pi^2 z}
\Big] = 0~.
\end{eqnarray}
For a value of $r$, we can plot in-medium effective mass of a nucleon against medium density using Eq.~(\ref{FinalEoSEq}), as shown in Sec.~\ref{Results}.

\subsection{Saturation Curve of the Large Nuclear Medium}
\label{SaturationCurve}
The nucleons interact predominantly by exchanging mesons. Within a large nuclear medium, nuclear interactions can have long wavelength and be over large distances. In such cases, all the hadronic d.o.f. of the medium are expected to be dealt dynamically. To describe such a system, we use a relativistic mean field theory model as proposed by Walecka\cite{Walecka1974, Walecka1997, Walecka2004}. We have already derived the self-energies of a quasinucleon in Sec.~\ref{SelfE} which we will use in the Walecka model to find the saturation curve of a large system of nucleons from where we should be able to obtain the minimum of their individual in-medium energy(\textit{i.e.} saturation energy) at a certain value of density(\textit{i.e.} saturation density).\\

\noindent Following~\cite{Walecka2004}, let us start with a nucleon field $\psi$ has two isospin components, each with two spin degrees of freedom
  \begin{align}
    \psi &= \begin{bmatrix}
           p_{\uparrow \downarrow} \\
           n_{\uparrow \downarrow}
         \end{bmatrix},
  \end{align}
where $p$ and $n$ denote proton and neutron fields respectively. Consider a system of volume $\mathcal{V}$ where exists uniformly $B$ nucleons. Let us further imagine a scenario in which the system is highly compressed so that the nucleon density $\rho_N = B/\mathcal{V}$ is large. Each nucleon acts as the source for a neutral scalar meson field $\phi$ with bare mass $m_S$ coupled to the scalar density, $\bar{\psi} \psi$, of the nucleon and a massive neutral vector meson field $V_\mu$ with bare mass $m_V$ coupled to the nucleon current, $i \bar{\psi} \gamma_\mu \psi$. The choice of the above mentioned degrees of freedom seems to be justified in view of the large attractive scalar and repulsive vector potential empirically found in nucleon scattering phenomena. The lagrangian density of such a system can be written using these fields:
\begin{eqnarray}
\label{lagrangian_original}
\mathcal{L} = - \frac{1}{4} F_{\mu \nu} F^{\mu \nu}  - \frac{1}{2} m_V^2 V_\mu V^\mu  - 
\frac{1}{2} \left[
\partial_\mu \phi \partial^\mu \phi + m_S^2 \phi^2
\right] - \bar{\psi} \left[
\gamma_\mu \left( \partial_\mu - i g_V V_\mu \right) + \left( M_N - g_S \phi  \right)
\right] \psi
\end{eqnarray}
where
\begin{eqnarray}
F_{\mu \nu} = \partial_\mu V_\nu - \partial_\nu V_\mu ~.
\end{eqnarray}
We work with this theory in imaginary time so that $x_\mu \equiv (\vec{x}, ix_0) = (\vec{x}, it)$. In the limit where the system is sufficiently compressed, the meson fields become classical and can be approximated to their expectation values:
\begin{eqnarray}
\phi &\rightarrow & \langle \phi \rangle ~\equiv \phi_0,\\
V_\mu &\rightarrow & \langle V_\mu \rangle \equiv i \delta_{\mu 4} V_0.
\end{eqnarray}
Notice here that the spatial part of the vector field vanishes due to the rotational symmetry of a sufficiently large system. Example of such a system can be the interior of a neutron star. But for nuclei, even for the heavy ones such as $^{208}\rm{Pb}$, such an assumption will not hold true in general. Furthermore, translational symmetry of such large, uniform medium tells us that the mean values of the meson fields, $\phi_0$ and $V_0$ will be constants. Calling for variation of action in the vector meson field along with the symmetry consideration that only the fourth component of the nucleon current $i \bar{\psi} \gamma_\mu \psi$ survives because of the $\delta_{\mu 4}$ in the vector field, we get
\begin{eqnarray}
V_0 = \frac{g_V}{m_V^2}\rho_N~.
\end{eqnarray}
Conservation of nucleon number will give us a constant $\rho_N$ and hence value of $V_0$ in terms of conserved quantities. Considering all these findings in this section so far, we can reduce the lagrangian in Eq.~(\ref{lagrangian_original}) to the mean field lagrangian
\begin{eqnarray}
\mathcal{L}_{MFT} = \frac{1}{2} m_V^2 V_0^2 - \frac{1}{2} m_S^2 \phi_0^2 - \bar{\psi} \left[
\gamma_\mu \partial^\mu + \gamma_4 g_V V_0 + M_N^*
\right] \psi ,
\end{eqnarray}
where $M_N^*$ is the effective mass of the nucleon:
\begin{eqnarray}
\label{EffMassNuc}
M_N^* = M_N - g_S \phi_0~.
\end{eqnarray}
The nucleon field can be expanded in quantum field operators and in Schr\"{o}dinger picture
\begin{eqnarray}
\hat{\psi}(\vec{x}) = \frac{1}{\sqrt{\mathcal{V}}} \sum_{\vec{k}, \lambda}
\left[
u(\vec{k}, \lambda) A_{\vec{k}, \lambda} e^{i\vec{k}.\vec{x}} +
v(-\vec{k}, \lambda) B_{\vec{k}, \lambda} e^{-i\vec{k}.\vec{x}}
\right]
\end{eqnarray}
with periodic boundary conditions in volume $\mathcal{V}$. Here $u, v$ denote Dirac spinors. Quantization of this theory is achieved by imposing anticommutation relations:
\begin{eqnarray}
\Big\{
A_{\vec{k}, \lambda}, A^\dagger_{\vec{k}', \lambda'}
\Big\} &=& \delta_{\vec{k} \vec{k}'} \delta_{\lambda \lambda'},\\
\Big\{
B_{\vec{k}, \lambda}, B^\dagger_{\vec{k}', \lambda'}
\Big\} &=& \delta_{\vec{k} \vec{k}'} \delta_{\lambda \lambda'}~.
\end{eqnarray}
The corresponding Hamiltonian density is given by
\begin{eqnarray}
\label{MFHamiltonian}
\mathcal{H}_{MFT} &=& \left( \frac{\partial \mathcal{L}_{MFT}}{\partial \dot{\psi}} \right) \dot{\psi} - \mathcal{L}_{MFT} \\
&=& 
\frac{1}{2} m_S^2 \phi_0^2 - \frac{1}{2} m_V^2 V_0^2 + 
g_V V_0 \rho_N
 + \frac{1}{\mathcal{V}} \sum_{\vec{k}, \lambda} (\vec{k}^2 + {M_N^*}^2)^{1/2} \left(
A_{\vec{k}\lambda}^\dagger A_{\vec{k}\lambda} + B_{\vec{k}\lambda}^\dagger B_{\vec{k}\lambda}
\right)~.
\end{eqnarray}
For uniform nuclear matter in ground state, the nucleons can approximated as a free Fermi gas (a nucleon can move freely throughout the volume at a mean potential generated by all other nucleons in the system) that fills up all the momentum states up to the Fermi level $k_F$ with a degeneracy of $\gamma$. Considering degeneracy in the spin and isospin degrees of freedom of the nucleon field for each momentum state, $\gamma = 4$. In the above expression, $\left(
A_{\vec{k}\lambda}^\dagger A_{\vec{k}\lambda} + B_{\vec{k}\lambda}^\dagger B_{\vec{k}\lambda}
\right)/\mathcal{V}$ is the number density which we can replace with  $\frac{\gamma}{(2 \pi)^3}\int_0^{k_F} d^3 \vec{k}$ for a large enough system. We Additionally use Eq.~(\ref{EffMassNuc}) to finally write the expression for the energy density of the medium
\begin{eqnarray}
\label{EnergyDensityFinal}
\mathcal{E}(\rho_N, \phi_0) 
&=& 
\frac{1}{2} m_S^2 \phi_0^2 - \frac{1}{2} m_V^2 V_0^2 + 
g_V V_0 \rho_N
 + \frac{\gamma}{(2 \pi)^3} \int_0^{k_F} d^3 \vec{k}~ (\vec{k}^2 + {M_N^*}^2)^{1/2}~.
 \label{EnergyDensityFinal}
\end{eqnarray}

\section{Results}
\label{Results}
Comparing the interaction terms of the lagrangian in Eq.~(\ref{lagrangian_original}) with the pole in the two-point correlator in Eq.~(\ref{Pheno1}), we can write
\begin{eqnarray}
\label{DefSigmas}
\phi_0 = \Sigma_S/g_S ~, \qquad V_0 = \Sigma_V/g_V ~.
\end{eqnarray}
Using Eq.~(\ref{Density}, \ref{DefSigmas}) in Eq.~(\ref{EnergyDensityFinal}), we find the average energy per nucleon,
\begin{eqnarray}
\frac{\mathcal{E}}{\rho_N} &=& \frac{M_N}{2C_S^2} \frac{(\Sigma_S/M_N)^2}{(\gamma/6 \pi^2) x_{\rm{sat}}^3}
+ \frac{1}{2}\Sigma_V + 
\frac{\gamma}{(2 \pi)^3 (\gamma/6 \pi^2) x_{\rm{sat}}^3 M_N^3} \int_0^{k_F} d^3 \vec{k}~ (\vec{k}^2 + {M_N^*}^2)^{1/2}~.
\end{eqnarray} 
where we write $k/M_N$ as $x$ and $k_F/M_N$ as $x_{\rm{sat}}$. 
$C_S^2$ and $C_V^2$ are defined as:
\begin{eqnarray}
C_S^2 = g_s^2 \left(\frac{M_N^2}{m_S^2}\right), \qquad C_V^2 = g_V^2 \left(\frac{M_N^2}{m_V^2}\right).
\end{eqnarray}
These are the two parameters in the relativistic mean field theory of nuclear matter which will be found in this section from the experimentally accessible properties such as the binding energy and density of uniform nuclear matter. We further use $M_N^*/M_N$ as $y_{\rm}$ and Eq.~(\ref{RatioSigmaV}) with $\gamma = 4$ to express the binding energy per nucleon as,
\begin{eqnarray}
\label{EqofStateNclrMed}
\frac{\mathcal{E}}{\rho_N} - M_N &=& 
\frac{3 \pi^2 M_N (1 - y_{\rm{sat}})^2}{4 x^3_{\rm{sat}} C_S^2}
+ \frac{4 M_N x_{\rm{sat}}^3 y_{\rm{sat}}}{3 (\pi^2r - 6 x^3_{\rm{sat}})}
+ \frac{3 M_N}{x_{\rm{sat}}^3} \int_0^{x_{\rm{sat}}} dx ~ x^2 ~ \sqrt{x^2 + y^2} - M_N ~.
\end{eqnarray}
Let's take a moment here to understand how quark-hadron duality is exploited in our calculations so far. In Sec.~\ref{Sec:2ptCorr}, we derived the OPE of the two-point correlator of local nucleon current in terms of simplest quark condensates of dimension three. In our calculations, certain quark condensate was taken into account along with the vacuum condensate which vanishes otherwise in the absence of a surrounding nuclear medium. Assuming that the OPE will have a pole at the effective nucleon mass, we could estimate in-medium scalar and vector self-energies. We use these self-energies in the theory assumed in Sec.~\ref{SaturationCurve} written in terms of hadron fields to calculate the binding energy per nucleon in a sufficiently large nuclear medium (so that finite size effect can be ignored) and extract the coupling parameters of the relativistic mean field theory lagrangian in Eq.~(\ref{lagrangian_original}). Values of these coupling parameters could be estimated by minimizing the energy of the system (i.e. Hamiltonian in Eq.~(\ref{MFHamiltonian})) as well (see~\cite{Walecka2004}) but in this work, minimization of energy is achieved naturally from the trade-off between average individual nucleon energy and the interaction energy arising from in-medium scalar and vector coupling (given in Eq.~(\ref{PiSBorelTransformed}), (\ref{PiqBorelTransformed}), and (\ref{PiuBorelTransformed})) that are dependent on the light quark condensates to the lowest order.\\

We evaluate Eq.~(\ref{FinalEoSEq}) and Eq.~(\ref{EqofStateNclrMed}) for different values of $x_{\rm{sat}} = k_F/M_N$ to generate the saturation curve for the infinite nuclear matter. We find that the saturation curve, $\mathcal{E}/\rho_N - M_N$, attains the minimum value of $-15.6~\rm{MeV}$ at the nuclear matter saturation density at $k_F = 1.33 ~{\rm{fm}}^{-1}$ for:
\begin{itemize}
\item $r = 0.029$ giving us $\langle \bar{q} q \rangle_{\rm{vac}} = -~(0.288 ~\rm{GeV})^3$ from Eq.~(\ref{def_r}), 
\item $C_S^2 = 235.62$,
\item $C_V^2 = 142.59$.
\end{itemize}
Using the above values of $r, C_S^2$ and $C_V^2$, we obtain the plots for the saturation curve of the infinite nuclear medium (Fig. \ref{fig:SatCurve}), effective mass of nucleon (Fig. \ref{fig:EffMass}), and in-medium vector self-energy (Fig. \ref{fig:VecEnergy}), as provided below. Here the value of effective nucleon mass, $M_N^*/M_N$, at the nuclear matter saturation density is 0.608 (corresponding to $M_N^*=0.570~\rm{GeV}$). The value of in-medium vector self-energy at the saturation density($\Sigma_V(x_{\rm{sat}})$) is found to be around $0.178~\rm{GeV}$. We use these results in Sec.~\ref{Ch_NS} and Sec.~\ref{Ch_Regime} to derive and discuss the critical mass of a neutron star.\\

The values of the coupling parameters are comparable to the values found by Walecka: 
\begin{itemize}
\item $C_S^2 = 266.9$, $C_V^2 = 195.7$ in ref.~\cite{Walecka1974},
\item $C_S^2 = 357.4$, $C_V^2 = 273.8$ in ref.~\cite{Walecka1997}.
\end{itemize}
The value of the light quark condensate obtained in this work is comparable to the values obtained in~\cite{LQCDDetermination, LQCDDetermination2} \textit{etc.}. We should also note that Furnstahl \textit{et. al.} used Monte Carlo approach to generate the condensate values which was used in the Sum Rules to obtain $M_N^* = 0.64^{+0.13}_{-0.09}$ GeV and $\Sigma_V = 0.29^{+0.06}_{-0.10}$ GeV at the saturation density\cite{Furnstahl1996}. These results had large uncertainties due to which determination of the binding energy of the order of -16 MeV was beyond the scope of that work.\\

Since we have considered only the operators till mass dimension 3 to produce this curve, behavior of the curve near the strong coupling domain (with matter density $>> x_{\rm{sat}}$) obtained from this work may not be accurate. Nevertheless, we can incorporate operators of higher dimension in the OPE to be able to predict the behavior near that domain.\\

\begin{figure}[H]
\centering
\includegraphics[scale=0.25]{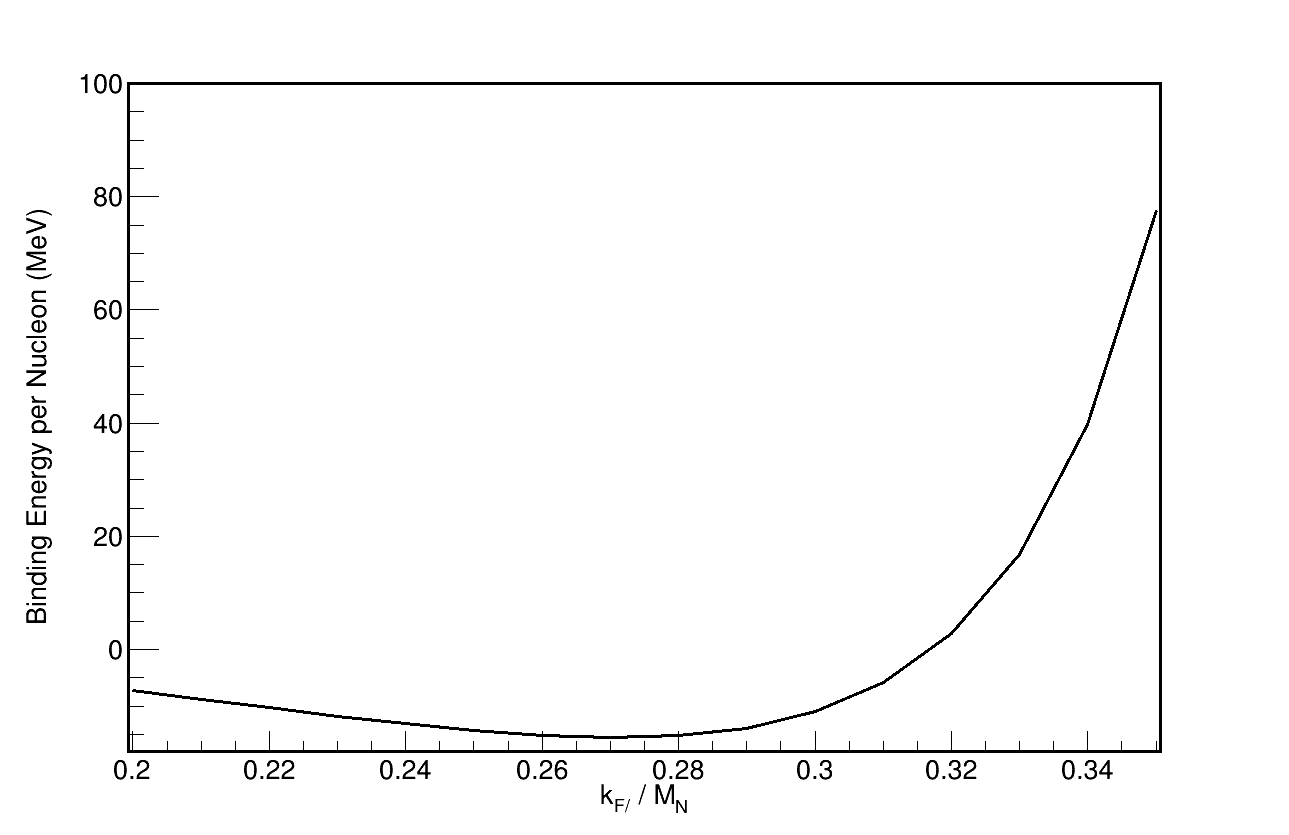}
\caption{Saturation curve for infinite nuclear matter. Binding energy per nucleon($\mathcal{E}/\rho_N - M_N$) is plotted against normalized Fermi wavenumber($k_F/M_N$) of infinite nuclear medium. Minimization of binding energy is achieved naturally through trade-off between average energy per nucleon and the interaction energy through in-medium scalar and vector coupling. After fitting the minimum of this curve to nuclear binding energy of -15.6 MeV and $k_F$ = $1.33 ~{\rm{fm}}^{-1}$, we obtain $\langle \bar{q} q \rangle = -(0.288~{\rm{GeV}})^3$, $C_S^2 = 235.62$, $C_V^2 = 142.59$.}
\label{fig:SatCurve}
\end{figure}
\begin{figure}[H]
\centering
\includegraphics[scale=0.25]{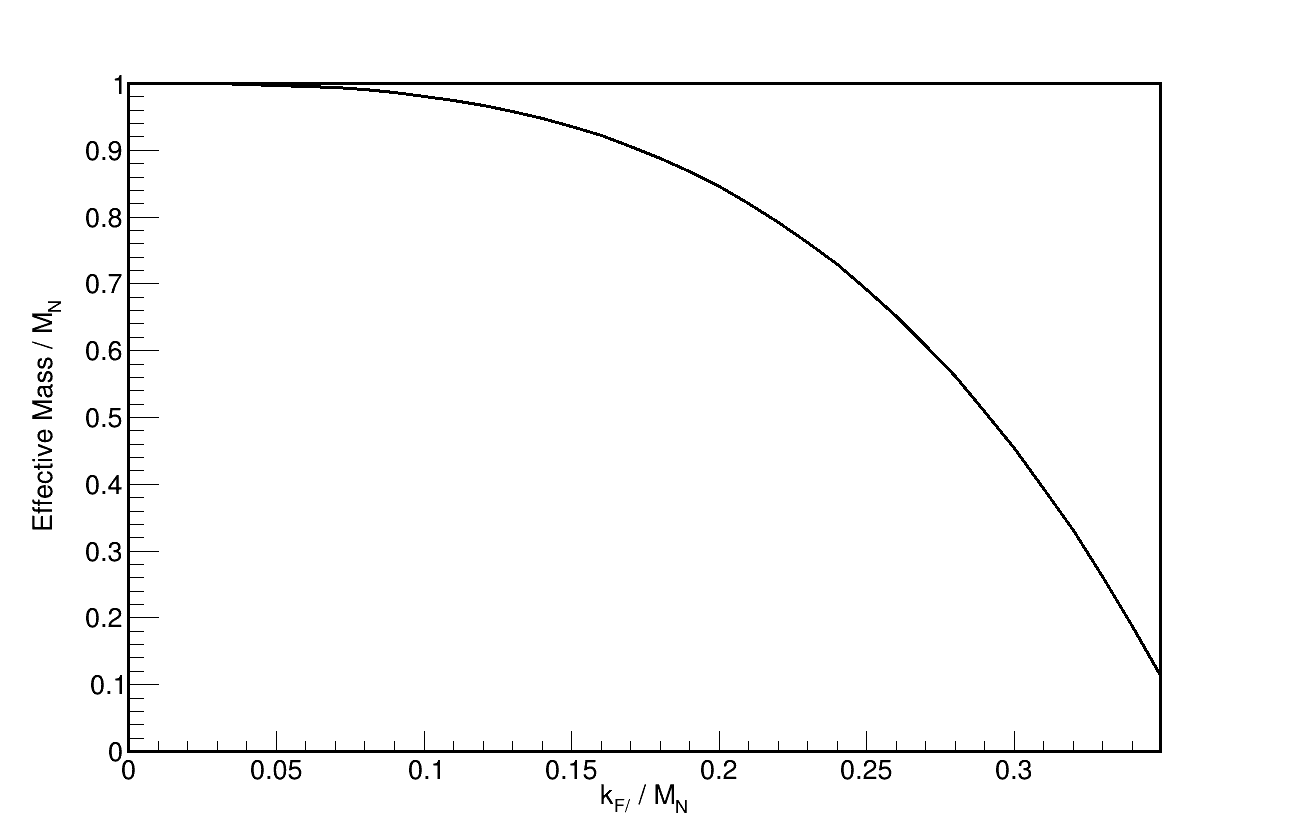}
\caption{Prediction for normalized effective mass ($M^*_N/M_N$) against normalized Fermi wavenumber ($k_F/M_N$). At the saturation density ($k_F = 1.33 ~{\rm{fm}}^{-1}$), we obtain $M^*_N = 0.570~\rm{GeV}$.}
\label{fig:EffMass}
\end{figure}
\begin{figure}[H]
\centering
\includegraphics[scale=0.25]{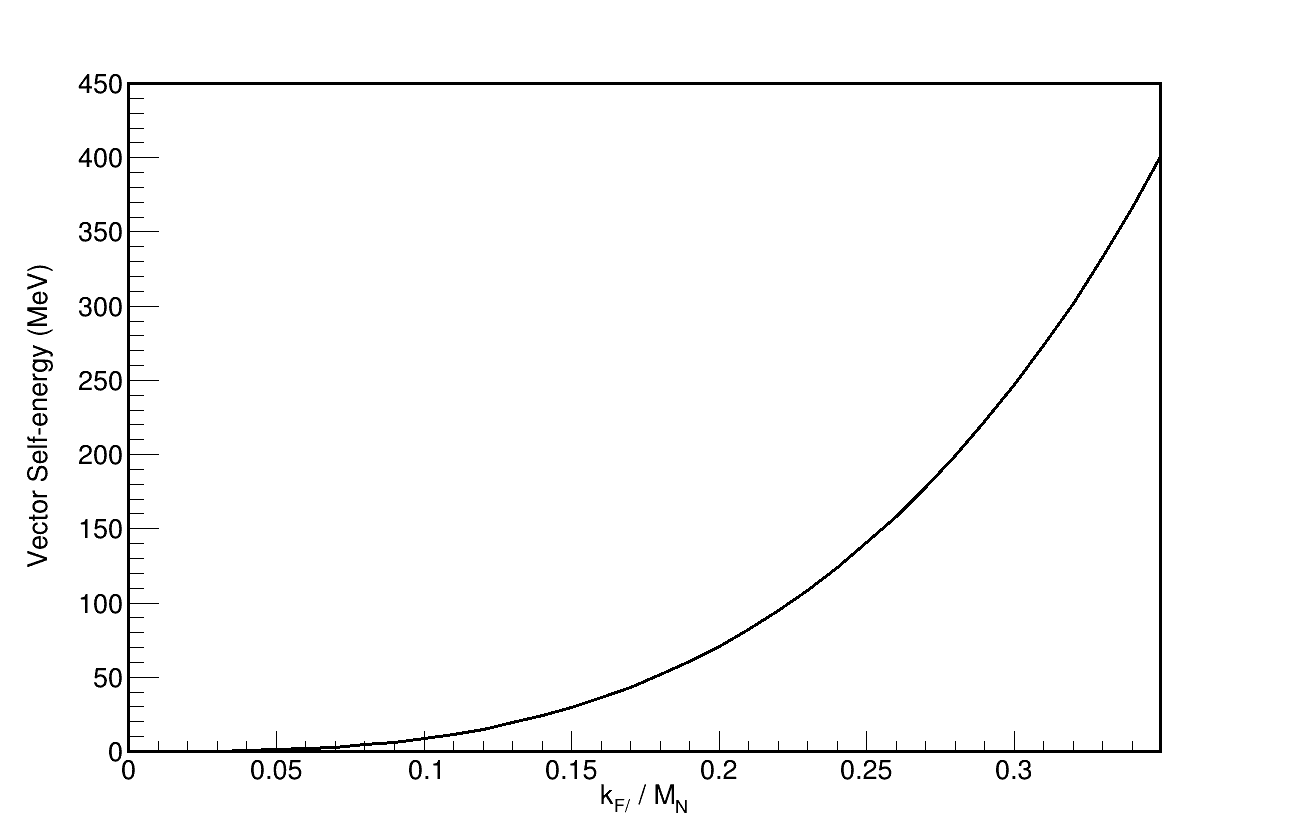}
\caption{Prediction for in-medium vector self-energy ($\Sigma_V$) as a function of normalized Fermi wavenumber ($k_F/M_N$). At the saturation density ($k_F = 1.33 ~{\rm{fm}}^{-1}$), we obtain $\Sigma_V = 0.178$ GeV.}
\label{fig:VecEnergy}
\end{figure}

\section{Obtaining the Mass Limit of Neutron Star and the Mass Gap}
\label{Sec:NS}
During the supernova explosion, a star can derive large enough energy so that its inner core collapses into a stable configuration consisting of closely packed nucleons. This new configuration would have a very small radius and extremely high density, and is called neutron star~\cite{Zwicky1, Zwicky2, Zwicky3, Zwicky4}. The mass and size of the neutron star depend critically on the equation of state of its constituent matter, $P \equiv P(\mathcal{E})$, obeying general relativistic equations in hydrostatic equilibrium (here $P(\mathcal{E})$ denotes the pressure profile for a given density profile within the stellar medium). In general relativity, there exists an upper limit for the mass of a star made of incompressible matter beyond which the internal pressure needed for the hydrostatic equilibrium of the star becomes infinite. It can be shown that, for a stable non-rotating cold neutron star with incompressible, constant-density nuclear matter treated in the general relativistic framework, the maximum possible mass is definitely less than $5M_\odot$~\cite{Phillips1996}. The interaction between nucleons in a neutron star plays an important role in determining the maximum mass of a neutron star. Inside a high-density nuclear medium, the degenerate nucleons have enough energy to produce pions and hyperons. In such a scenario, the interior of a neutron star can become more compressible due to reduction in the effective short range repulsion between nucleons, although the mass limits are negligibly affected in the presence of hyperon in the nuclear medium \cite{StrobelWeigel}.\\

The critical mass of a neutron star can depend on a number of other parameters as well, such as differential rotation~\cite{ShapiroRotNS}, anisotropic pressure~\cite{AnisoNS} \textit{etc.}. But the greatest uncertainty in the estimate of the critical mass comes from the fact that the equation of state within a neutron star is poorly known as the strong interactions within the constituent nuclear medium within a typical neutron star is expected to lie in the non-perturbative regime~\cite{Lattimer2015TheEO}. Rhoades and Ruffini~\cite{RhoadesMaxMass} considered the most extreme equation of state with three conditions: (a)causality, (b)Le Chatelier's principle, (c)general relativistic equations in hydrostatic equilibrium, to conclude that the critical mass of a neutron star with density always greater than $4.6 \times 10^{14}~\rm{g/cm^3}$ cannot be larger than $3.2M_\odot$. Furthermore, in the presence of non-zero angular momentum, this critical mass will be affected by a factor smaller than 1.5. Strobel and Weigel suggested that\cite{StrobelWeigel} the minimum possible mass of a neutron star depends on the earliest stage of its evolution, specifically the deleptonization period. For a cold, non-rotating neutron star, they estimated the minimum mass to be $\sim (0.88- 1.28)M_\odot$ and the maximum mass to be $\sim (1.70-2.66)M_\odot$.

\subsection{Solving Tolman-Oppenheimer-Volkoff Equations}
\label{Ch_NS}
Making the assumption of spherical symmetry, zero velocities (\textit{i.e.} static spacetime) and an ideal fluid model, Tolman, Oppenheimer and Volkoff (TOV) found the equation of state of neutron star in a General Relativisitic (GR) framework~\cite{TOV}. Here we write the first two TOV equations:
\begin{eqnarray}
\label{ScTOVF_1}
\frac{dP}{dr} &=& - \frac{G
\big[\mathcal{E}(r) + P(r)\big]\big[m(r) + \frac{4 \pi r^3 P(r)}{c^2}\big]
}{c^2r^2 \big[1 - \frac{2Gm(r)}{c^2r}\big]},\\
\label{ScTOVF_2}
\frac{dm}{dr} &=& 4 \pi r^2 \rho(r) = \frac{4 \pi r^2 \mathcal{E}(r)}{c^2},
\end{eqnarray}
where $P(r)$, $\rho(r)$ and $\mathcal{E}(r)$ are the steller pressure, mass density and energy density at proper radius $r$, and $m(r)$ is the total mass contained within $r$. There is considerable uncertainty in the equation of state for the densities of the nuclear medium that exist in the neutron star interior. In such a scenario, we can constrain the equation of state with assumptions and general principles to derive the critical mass. Here we assume uniform energy density of the neutron star interior which is incompressible at any finite pressure. With this assumption, solving Eq.~(\ref{ScTOVF_2}) becomes trivial. With the boundary conditions, $m(r = 0) = 0$ and $\mathcal{E}(r) = \mathcal{E}_0$ for all $r<R$  ($R$ is the neutron star proper radius, as defined in \cite{Buchdahl}), we get
\begin{equation}
\label{m_x}
m(r) = \frac{4 \pi \mathcal{E}_0 r^3}{3c^2}~.
\end{equation}
In the next section, we will plug in the value of the nuclear saturation energy density in the above equation to get the $m(r)$. Using the boundary conditions $(x = 0, P = P_0)$ and $(x = R, P=0)$, we solve Eq.~(\ref{ScTOVF_1}) to get:
\begin{eqnarray}
\label{SolutionTOV}
\ln \left[
\frac{P_0 + \mathcal{E}_0}{3 P_0 + \mathcal{E}_0}
\right] = \frac{1}{2} \ln \left[
1 - \left(\frac{8 \pi G \mathcal{E}_0}{3 c^4}\right) R^2
\right]~. 
\end{eqnarray}
It is evident from above that the critical mass of a neutron star strongly depends on the equation of state of its interior. We assume here uniform medium density assuming that the interior of the neutron star is a perfect fluid and its density does not increase outwards substantially \cite{Buchdahl}. The extension of this scenario to a rotating neutron star \cite{AngMomNS} with various equations of states~\cite{Lattimer2001, Reddy2012, Reddy2014} will be explored in later work.

\subsection{Deconfinement, Rigid Vector Repulsion, and Black Hole Regimes}
\label{Ch_Regime}
As the density of the nuclear medium increases, the interaction of the vector field between the nucleons is enhanced. It can be argued in such a scenario that the equation of state, $P_0 = \mathcal{E}_0$, is realized asymptotically~\cite{Zeldovich1962} as an essential feature of a relativistic field theory with the lagrangian that has a massive vector field. In this rigid vector repulsion(RVR) regime, the speed of sound in the interior of neutron star approaches $c_s = c$, giving us the maximum possible mass of a neutron star in the causal limit,
\begin{eqnarray}
\label{Eq:MCausal}
M_{NS}^{RVR} = \left[\frac{81c^8}{2048\pi G^3\mathcal{E}_0}\right]^{1/2} \approx 5.47M_\odot~,
\end{eqnarray}
from Eq.~(\ref{m_x}) and (\ref{SolutionTOV}). In this regime, the large vector repulsion between nucleons pushes the maximum mass of a neutron star to $5.47 M_\odot$. This is the hard limit in the mass of a neutron star beyond which it becomes a black hole.\\

However, the interior of a cold neutron star possibly goes through phase transition much before the density of a neutron star hits the rigid vector repulsion regime\cite{AnnalaNat2020}. This is attributed to the fact that the chiral symmetry gets restored in the medium much before the density approaches rigid vector repulsion regime. Quantitatively, this is understood from the behavior of the most dominant order parameter of chiral symmetry breaking \textit{i.e.} light quark condensate. The value of this condensate decreases in presence of a nuclear medium. In a scenario where the medium has sufficiently high density, the value of the condensate approaches zero signifying chiral symmetry restoration. This is the deconfinement regime where QCD state transits from hadronic to QGP phase.\\

From Eq.~(\ref{Density}), the saturation density of nuclear matter is estimated to be
\begin{eqnarray}
\label{Sat_Den}
\rho_N^{\rm{sat}} 
= 0.159 ~ \rm{fm}^{-3}~,
\end{eqnarray}

\noindent while, From Eq.~(\ref{condinmediumfinal}), we can write for restoration of chiral symmetry:
\begin{eqnarray}
\langle \bar{q}q \rangle_{\rho'_N} &\approx &
\langle \bar{q}q \rangle_{\rm{vac}}
+ \frac{\rho'_N \sigma_N}{2 m_q} = 0,\\
\label{ChSymRest_Cond}
\rho'_N &=& 
0.345 ~ {\rm{fm}}^{-3}~.
\end{eqnarray}

Eq.~(\ref{Sat_Den}) and (\ref{ChSymRest_Cond}) together imply that the interior of neutron star enters phase transition for a baryon density which is approximately twice the saturation density (similar result was found in~\cite{Jacobsen2007}). This is the relativistic regime where the asymptotic equation of state is given by $P_0 = \mathcal{E}_0/3$ and the speed of sound, $c_s = c/\sqrt{3}$. In this regime, the maximum mass of a neutron star is given by,
\begin{eqnarray}
\label{MConf}
M_{NS}^{rel} = \left[\frac{125c^8}{7776\pi G^3\mathcal{E}_0}\right]^{1/2} \approx 3.48M_\odot~,
\end{eqnarray}
from Eq.~(\ref{m_x}) and (\ref{SolutionTOV}). Once the neutron star core enters the deconfinement phase, its interior becomes more compressible and for a sufficiently dense QCD matter core, the speed of sound asymptotically hits the conformal limit, $c_s = c/\sqrt{3}$, where the trace of its energy-momentum tensor ($3P - \mathcal{E}$) vanishes and the neutron star reaches its maximum mass, $M_{NS}^{conf} = 3.48 M_\odot$. This sets a soft limit on neutron star mass because it is only less likely that a neutron star core can evade deconfinement to enter the RVR phase.\\

We have assumed in our calculation of maximum neutron star mass in all three regimes that the density of the neutron star interior is equal to saturation density. Neutron star with a highly compressible interior may have a density which is much higher than the saturation density. The maximum mass is inversely proportional to the square root of the medium density, lowering even more the critical mass of a neutron star in all three regimes. Therefore, it should be rare to find an isolated neutron star with a mass greater than $3.48 M_\odot$ and impossible to find a neutron star with a mass beyond $5.47 M_\odot$. Additionally, in the absence of any conceivable mechanism for the speed of sound to reach the causal limit within the hadronic or QGP core of a star with a mass less than the critical value derived in Eq.~(\ref{Eq:MCausal}), it seems unlikely that a black hole with a mass smaller than $5.47 M_\odot$ can be formed through supernova explosion. Hence, there should plausibly exist a gap, $\Delta \in [3.48 M_\odot, 5.47 M_\odot]$,  in the stellar mass spectrum where $\Delta$ is not heavy enough to evolve into a black hole and is not light enough for a neutron star to exist as a stable configuration with quark-matter core.\\

This hypothesis is consistent with the observations of supermassive neutron stars/light black hole candidates and their masses, as listed in~\cite{SuperMassiveNS}. Advanced LIGO and Advanced Virgo recently listed 35 binary coalescence candidates in their third Gravitational Wave Transient Catalog\cite{GTWC3_2021} out of which only four binaries are found to involve compact objects whose estimated masses lies statistically within $[3.48M_\odot, 5.47M_\odot]$ with the expected values beyond the same gap proposed in our work:
\begin{itemize}
\item GW200115 042309: involves an object $5.9^{+2.0}_{-2.5}~M_\odot$.
\item GW191113 071753: involves an object $5.9^{+4.4}_{-1.3}~M_\odot$.
\item GW200316 215756: involves an object $7.8^{+1.9}_{-2.9}~M_\odot$.
\item GW200322 091133: involves an object $14.0^{+16.8}_{-8.7}~M_\odot$.
\end{itemize}

Recently, a compact binary system that involves a $22.2-24.3~M_\odot$ black hole and a compact object with a mass of $2.50-2.67 ~M_\odot$ from the gravitational wave signal GW190814 has been reported in~\cite{BinaryBHNS} with the speculation that the latter component is either a light black hole or the heaviest neutron star observed till date. The heaviest neutron star confirmed so far is the companion of PSR J0952-0607 with a mass $2.35 \pm 0.17 M_\odot$\cite{Romani2022}. Ref.~\cite{PhotoAstrometric2022} presents a photometric-isometric combined analysis of five neutron star/stellar mass black hole candidates identified in gravitational microlensing surveys\cite{OGLE,MOA1, MOA2}, and one candidate MOA-2011-BLG-191/OGLE-2011-BLG-0462 is shown to have a mass of $1.6-4.4~M_\odot$. This claim is negated though in~\cite{Sahu2022} where the lens mass obtained is $7.1 \pm 1.3~M_\odot$ which is much beyond the upper end of our proposed limit. This claim is further supported by~\cite{Mroz2022} which claims that there are systematic errors in the analysis provided in~\cite{PhotoAstrometric2022} and the lens mass of OGLE-2011-BLG-0462 should be $7.88 \pm 0.82~M_\odot$. Thompson \textit{et. al.} recently reported their observation\cite{Thompson2019} about the presence of an unseen companion of the red giant 2MASS J05215658+4359220 with a mass $3.3^{+2.8}_{-0.7} ~M_\odot$. But van den Heuvel and Tauris argued in~\cite{Heuvel} that the unseen candidate can be a close binary of two main-sequence stars, primarily because no emission of X-ray is detected so far from the candidate. For recent reviews on the neutron star-black hole mass gap, see~\cite{BH-NS1, BH-NS2, DeSa2022, Gao2023}.

\section{Conclusion}
We started with an OPE of the two-point correlator of local nucleon current in terms of all the operators up to mass dimension three. We exploited quark-hadron duality to compare the OPE with hadron phenomenological spectrum to calculate the scalar and vector self-energy of a nucleon as functions of density of the surrounding infinite nuclear medium. We provided the effective mass plot of the in-medium nucleon as a function of Fermi momentum. We used the self-energy terms in the relativistic mean field lagrangian as proposed by Walecka to generate the nuclear saturation curve. Fitting the minimum of the saturation curve to the nuclear binding energy of $-15.6~\rm{MeV}$ at the Fermi momentum of $k_F = 1.33~\rm{fm}^{-1}$, we obtained the value of light quark condensate $\langle \bar{q} q \rangle = -(0.288~\rm{GeV})^3$ and the coupling parameters of the Walecka lagrangian: $C_S^2 = 235.62$, $C_V^2 = 142.59$. \\

We used Tolman-Oppenheimer-Volkoff Equations to calculate the critical mass of a neutron star for a uniform nuclear saturation energy density in its interior. The maximum possible mass of a neutron star is achieved only in the limiting case where the vector potential between nucleons with the NS starts to predominate all other interactions, so that the EoS, $P = \mathcal{E}$ (i.e. speed of sound = $1$) is asymptotically achieved. But the interior of the NS possibly goes through phase transition much before its density reaches the level where the vector potential is sufficiently large. This leads to a much softened EoS, $P = \mathcal{E}/3$ (i.e. speed of sound = $1/\sqrt{3}$). Because it is less likely for a neutron star to avoid this phase transition so that vector potential can build up to asymptotically enhance the speed of sound to 1, implying a plausible gap in the neutron star-black hole mass spectrum expressed in terms of universal constants,
\begin{eqnarray*} 
\Delta \in  \left[\sqrt{\frac{125c^8}{7776\pi G^3\mathcal{E}_0}}, \sqrt{\frac{81c^8}{2048\pi G^3\mathcal{E}_0}}\right] \approx [3.48M_\odot, 5.47M_\odot]
\end{eqnarray*}
where it is rare to find a isolated, cold, non-rotating neutron star or a black hole (here $\mathcal{E}_0$ = saturation energy density of infinite nuclear matter).\\

A natural extension of this framework will include gluon condensate and other higher order condensates in the OPE, and consider various equations of states to analyze the properties of the neutron star interior in more detail.

\section{Acknowledgements}

Author D.D. acknowledges the facilities of Saha Institute of Nuclear Physics, 
Kolkata, India. Author L.S.K. acknowledges support from the P25 group at Los 
Alamos National laboratory.



\begin{thebibliography}{9}

\bibitem{Riordan:1992hr}
E. M. Riordan, \textit{The Discovery of quarks}, Science, \textbf{256} (1992) 1287



\bibitem{Ellis:2014rma}
J. Ellis, \textit{The Discovery of the Gluon}, Int. J. Mod. Phys. A, \textbf{29 (31)} (2014) 1430072
  

\bibitem{Marciano:1977su}
W. J. Marciano and H. Pagels, \textit{Quantum Chromodynamics: A Review}, Phys. Rept., \textbf{36} (1978) 137


\bibitem{Shuryak:1980tp}
E. V. Shuryak, \textit{Quantum Chromodynamics And The Theory Of Superdense Matter}, Phys. Rept., \textbf{61} (1980) 71


\bibitem{Kisslinger:2014uda} 
L. S. Kisslinger and D. Das, \textit{Review of QCD, Quark-Gluon Plasma, Heavy Quark Hybrids, and Heavy Quark State Production in p-p and A-A Collisions}, Int.\ J.\ Mod.\ Phys.\ A, \textbf{31 (07)} (2016) 1630010

\bibitem{Adams:2005dq}
  J. Adams {\it et al.} [STAR Collaboration], \textit{Experimental and theoretical challenges in the search for the quark gluon plasma: The STAR Collaboration's critical assessment of the evidence from RHIC collisions}, Nucl. Phys. A, {\bf 757} (2005) 102

\bibitem{Adcox:2004mh}
  K.~Adcox {\it et al.} [PHENIX Collaboration], \textit{Formation of dense partonic matter in relativistic nucleus-nucleus collisions at RHIC: Experimental evaluation by the PHENIX collaboration}, Nucl.\ Phys.\ A, {\bf 757} (2005) 184

\bibitem{CBMFAIRQCD}
T.~Ablyazimov {\it et al.}, \textit{Challenges in QCD matter physics --The scientific programme of the Compressed Baryonic Matter experiment at FAIR}, The European Physical Journal A,  {\bf 53} (2016) 1


\bibitem{Citron:2018lsq}
  Z.~Citron {\it et al.},
  \textit{Report from Working Group 5 : Future physics opportunities for high-density QCD at the LHC with heavy-ion and proton beams}, CERN Yellow Rep. Monogr. (2019) 1159


\bibitem{EIC2022}
  R. A. Khalek {\it et al.}, \textit{Science Requirements and Detector Concepts for the Electron-Ion Collider},
  Nucl.\ Phys.\ A {\bf 122447}, 1026 (2022).


  


\bibitem{Walecka1974}
J. D. Walecka, \textit{A Theory of Highly Condensed Matter}, Ann. Phys., \textbf{83 (2)}  (1974) 491

\bibitem{Walecka1997}
B. D. Serot, J. D. Walecka, \textit{Recent Progress in Quantum Hadrodynamics}, Int. J. Mod. Phys., \textbf{06 (04)} (1997) 515


\bibitem{Coleman1973}
S. Coleman and D. J. Gross, \textit{Price of Asymptotic Freedom}, Phys. Rev. Lett., \textbf{31} (1973) 851 
\bibitem{Weinberg1973}
S. Weinberg, \textit{Non-Abelian Gauge Theories of the Strong Interactions}, Phys. Rev. Lett. \textbf{31} (1973) 494
\bibitem{Gross1973}
D. J. Gross and F. Wilczek, \textit{Asymptotically Free Gauge Theories. I}, Phys. Rev. D, \textbf{8} (1973) 3633
\bibitem{Gross1974}
D. J. Gross and F. Wilczek, \textit{Asymptotically Free Gauge Theories. II}, Phys. Rev. D, \textbf{9} (1974) 980


\bibitem{PiPi}
J. Gasser, H. Leutwyler, \textit{Chiral perturbation theory to one loop}, Ann. Phys., \textbf{158 (1)} (1984) 142

\bibitem{PiN}
J. Gasser, M. E. Sainio, A. \v{S}varcab, \textit{Nucleons with chiral loops}, Nucl. Phys. B, \textbf{307 (4)} (1988) 779

\bibitem{SignProblem1}
E. Y. Loh Jr., J. E. Gubernatis, R. T. Scalettar, S. R. White, D. J. Scalapino, and R. L. Sugar, \textit{Sign problem in the numerical simulation of many-electron systems}, Phys. Rev. B, \textbf{41} (1990) 9301

\bibitem{SignProblem2}
Stephen D.H. Hsu, David Reeb, \textit{On the sign problem in dense QCD}, Int. J. Mod. Phys. A, \textbf{25} (2010) 53

\bibitem{Baym:1978jf}
G.~Baym and C.~Pethick,``Physics of Neutron Stars,''
Ann. Rev. Astron. Astrophys. \textbf{17} (1979), 415-443


\bibitem{McLerran:2018hbz}
L.~McLerran and S.~Reddy,``Quarkyonic Matter and Neutron Stars,''
Phys. Rev. Lett. \textbf{122} (2019) no.12, 122701

\bibitem{Kisslinger:2021ach}
L.~S.~Kisslinger and D.~Das,``Review of low energy nuclear astrophysics,''
Mod. Phys. Lett. A \textbf{36} (2021) no.22, 2130019

\bibitem{Weinberg1967}
S. Weinberg, \textit{Precise relations between the spectra of vector and axial-vector mesons},, Phys. Rev. Lett., \textbf{18} (1967) 507

\bibitem{Weinberg1968}
S. Weinberg, \textit{Nonlinear realizations of chiral symmetry}, Phys. Rev., \textbf{166} (1968) 1568

\bibitem{Coleman1969First}
S. Coleman, J. Wess, and B. Zumino, \textit{Structure of Phenomenological Lagrangians. I}, Phys. Rev., \textbf{177} (1969) 2239

\bibitem{Coleman1969Second}
C. G. Callan, S. Coleman, J. Wess, and B. Zumino, \textit{Structure of Phenomenological Lagrangians. II}, Phys. Rev., \textbf{177} (1969) 2247

\bibitem{Dashen1969First}
R. Dashen, \textit{Chiral $SU(3)\otimes SU(3)$ as a Symmetry of the Strong Interactions}, Phys. Rev., \textbf{183} (1969) 1245

\bibitem{Dashen1969Second}
R. Dashen and M. Weinstein, \textit{Soft Pions, Chiral Symmetry, and Phenomenological Lagrangians}, Phys. Rev. \textbf{183} (1969) 1261

\bibitem{Teramond2005}
G. F. de Teramond and S. J. Brodsky, \textit{Hadronic Spectrum of a Holographic Dual of QCD}, Phys. Rev. Lett., \textbf{94} (2005) 201601

\bibitem{Erlich2005}
J. Erlich, E. Katz, D. T. Son, and M. A. Stephanov, \textit{QCD and a Holographic Model of Hadrons}, Phys. Rev. Lett., \textbf{95} (2005) 261602

\bibitem{Rold2005}
L. D. Rold and A. Pomarol, \textit{Chiral symmetry breaking from five dimensional spaces}, Nucl. Phys. B, \textbf{721} (2005) 79

\bibitem{Karch2006}
A. Karch, E. Katz, D. T. Son, and M. A. Stephanov, \textit{Linear confinement and AdS/QCD}, Phys. Rev. D, \textbf{74} (2006) 015005

\bibitem{Brodsky2015}
S. J. Brodsky, G. F. de T\'{e}ramond, H. G. Dosch, J. Erlich,
\textit{Light-front holographic QCD and emerging confinement}, Phys. Rep., \textbf{584} (2015) 1

\bibitem{Alho2013}
T. Alho, N. Evans, and K. Tuominen, \textit{Dynamic AdS/QCD and the spectrum of walking gauge theories}, Phys. Rev. D, \textbf{88} (2013) 105016

\bibitem{Jarvinen2022}
M. J{\"a}rvinen, \textit{Holographic modeling of nuclear matter and neutron stars}, Eur. Phys. J. C, \textbf{82} (2022) 282

\bibitem{Shifman1979First}
M. A. Shifman, A. I. Vainshtein, and V. I. Zakharov, \textit{QCD and resonance physics. theoretical foundations}, Nucl. Phys. B, \textbf{147} (1979) 385 

\bibitem{Shifman1979Second}
M. A. Shifman, A. I. Vainshtein, and V. I. Zakharov, \textit{QCD and resonance physics. applications}, Nucl. Phys. B, \textbf{147} (1979) 448

\bibitem{ReindersQuarkProp}
L.J. Reinders, H. Rubenstein, and S, Yazaki. \textit{Hadron properties from QCD sum rules},
Phys. Reports, \textbf{127} (1985) 1

\bibitem{LSKBS1}
L. S. Kisslinger, B. Singha, \textit{Charm, Bottom, Strange Baryon Masses Using QCD Sum Rules}, Int. J. Modern Phys. A, \textbf{33} (2018) 1850139

\bibitem{LSKBS2}
L. S. Kisslinger, B. Singha, \textit{Charmed baryon decay to a strange baryon plus a pion using QCD sum rules}, Int. J. Modern Phys. A, \textbf{34} (2019) 1950015

\bibitem{Cohen1991}
T. D. Cohen, R. J. Furnstahl, and David K. Griegel, \textit{From QCD Sum Rules to Relativistic Nuclear Physics}, Phys. Rev. Lett., \textbf{67 (8)} (1991) 961 

\bibitem{Cohen1992}
R. J. Furnstahl, David K. Griegel, and Thomas D. Cohen, \textit{QCD sum rules for nucleons in nuclear matter}, Phys. Rev. C, \textbf{46} (1992) 1507

\bibitem{Cohen1993}
X. Jin, T. D. Cohen, R. J. Furnstahl, D. K. Griegel, \textit{QCD sum rules for nucleons in nuclear matter II}, Phys. Rev. C, \textbf{47} (1993) 2882

\bibitem{Cohen1994}
X. Jin, M. Nielsen, T. D. Cohen, R. J. Furnstahl, and D. K. Griegel, \textit{QCD sum rules for nucleons in nuclear matter III}, Phys. Rev. C, \textbf{49} (1993) 464

\bibitem{Cohen1995}
 T. D. Cohen, R. J. Furnstahl, D. K. Griegel, X. Jin. \textit{QCD Sum Rules and Application to Nuclear Physics.} Prog. Part. Nucl. Phys., \textbf{35} (1995) 221-298
 
\bibitem{Walecka2004-1}
J. D. Walecka, \textit{Theoretical Nuclear and Subnuclear Physics}, Ed. 2, Imperial College Press (2004) 26
 
\bibitem{Drukarev1990}
E. G. Drukarev and E. M. Levin, \textit{The QCD Sum Rules and Nuclear Matter}, Nucl. Phys.
A, \textbf{511} (1990) 679; Nucl. Phys. A, \textbf{616} (1990) 715

\bibitem{Drukarev1994}
E. G. Drukarev, M. G. Ryskin, \textit{QCD sum rules and the properties of nuclear matter}, Nuclear Physics A, \textbf{578} (1994) 333 

\bibitem{Reya1974}
E. Reya, \textit{Chiral symmetry breaking and meson-nucleon sigma commutators: A review}, Rev. Mod. Phys., \textbf{46} (1974) 545

\bibitem{Jin1994}
X. Jin and R. J. Furnstahl, \textit{QCD sum rules for $\Lambda$ hyperons in nuclear matter}, Phys. Rev. C, \textbf{49} (1994) 1190

\bibitem{Jin1995}
X. Jin and M. Nielsen, \textit{QCD sum rules for Sigma hyperons in nuclear matter}, Phys. Rev. C, \textbf{51} (1995) 347

\bibitem{Gasser1991}
J. Gasser, H. Leutwyler, M. E. Sainio, \textit{Sigma-term update}, Phys. Lett. B, \textbf{253} (1991) 252 

\bibitem{Walecka2004}
J. D. Walecka, \textit{Theoretical Nuclear and Subnuclear Physics}, Ed. 2, Imperial College Press (2004) 119

\bibitem{LQCDDetermination}
C. McNeile, A. Bazavov, C. T. H. Davies, R. J. Dowdall, K. Hornbostel, G. P. Lepage, and H. D. Trottier, \textit{Direct determination of the strange and light quark condensates from full lattice QCD}, Phys. Rev. D, \textbf{87} (2012) 034503 

\bibitem{LQCDDetermination2}
Ting-Wai Chiu and Tung-Han Hsieh, \textit{Light quark masses, chiral condensate and quark–gluon condensate in quenched lattice QCD with exact chiral symmetry}, Nucl. Phys. B, \textbf{673} (2003) 217

\bibitem{Furnstahl1996}
R. J. Furnstahl, X. Jin, D. B. Leinweber, \textit{New QCD sum rules for nucleons in nuclear matter}, Phys. Lett. B, \textbf{387} (1996) 253


\bibitem{Zwicky1}
W. Baade, F. Zwicky, Phys. Rev.
\textbf{45} (1934) 138
\bibitem{Zwicky2}
W. Baade, F. Zwicky, \textit{On Super-Novae}, Proc. National Acad. Sci., \textbf{20(5)} (1934) 254
\bibitem{Zwicky3}
W. Baade, F. Zwicky, \textit{Cosmic Rays from Super-Novae}, Proc. National Acad. Sci., \textbf{20} (1934) 259
\bibitem{Zwicky4}
W. Baade, F. Zwicky, \textit{Remarks on Super-Novae and Cosmic Rays}, Phys. Rev., \textbf{46} (1934) 76

\bibitem{Phillips1996}
A. C. Phillips, \textit{The Physics of Stars}, John Wiley and Sons (1996), 192

\bibitem{StrobelWeigel}
K. Strobel and M. K. Weigel, \textit{
On the minimum and maximum mass of neutron stars and the delayed collapse}, Astron. Astrophys., \textbf{367} (2001) 582

\bibitem{ShapiroRotNS}
I. A. Morrison, T. W. Baumgarte, and S. L. Shapiro, \textit{
Effect of Differential Rotation on the Maximum Mass of Neutron Stars: Realistic Nuclear Equations of State,} Astrophys. J., \textbf{610} (2004) 941

\bibitem{AnisoNS}
A. Sulaksono, \textit{Anisotropic pressure and hyperons in neutron stars,} Int. J. Mod. Phys. E, \textbf{24} (2015) 1550007

\bibitem{Lattimer2015TheEO}
J. M. Lattimer and M. Prakash, \textit{The Equation of State of Hot, Dense Matter and Neutron Stars}, Phys. Rept., \textbf{621} (2016) 127; J. M. Lattimer and M. Prakash, \textit{Neutron Star Structure and the Equation of State}, Phys. Rep., \textbf{121} (2000) 333; \textit{The Physics of Neutron Stars}, Science, \textbf{304} (2004) 304; \textit{Neutron star observations: Prognosis for equation of state constraints}, Phys. Rep. \textbf{442} (2007) 109

\bibitem{RhoadesMaxMass}
C. E. Rhoades Jr. and R. Ruffini, \textit{Maximum Mass of a Neutron Star,} Phys. Rev. Lett., \textbf{32 (6)} (1972) 324

\bibitem{TOV}
R. C. Tolman, \textit{Static Solutions of Einstein's Field Equations for Spheres of Fluid}, Phys. Rev. \textbf{55} (1939) 364; 
J. R. Oppenheimer and G. M. Volkoff, \textit{On Massive Neutron Cores}, Phys. Rev. \textbf{55} (1939) 374

\bibitem{Buchdahl}
H. A. Buchdahl, \textit{General Relativistic Fluid Spheres}, Phys. Rev. \textbf{116 (4)} (1959) 1027

\bibitem{AngMomNS}
M. Hashimoto and K. Oyamatsu, \textit{Upper limit of the angular velocity of neutron stars}, Astrophys. J., \textbf{436(1)} (1994) 257

\bibitem{Lattimer2001}
J. M. Lattimer, M. Prakash, \textit{Neutron star structure and the equation of state}, Astrophys. J. \textbf{550}, (2001) 426

\bibitem{Reddy2012}
S. Gandolfi, J. Carlson, S. Reddy, \textit{The equation of state of neutron matter, symmetry energy and neutron star structure}, Phys. Rev. C, \textbf{85 (3)} (2012) 032801


\bibitem{Reddy2014}
S. Gandolfi, J. Carlson, S. Reddy, A. W. Steiner, R. B. Wiringa, \textit{The equation of state of neutron matter, symmetry energy and neutron star structure}, Eur. Phys. J. A \textbf{50}, (2014) 1

\bibitem{Zeldovich1962}
Ya. B. Zel'dovich, \textit{The equation of state at ultrahigh densities and its relativistic limitations}, JETP, \textbf{14 (5)} (1962) 1143

\bibitem{AnnalaNat2020}
E. Annala, T. Gorda, A. Kurkela, \textit{et. al.}, \textit{Evidence for quark-matter cores in massive neutron stars}, Nat. Phys. \textbf{16} (2020) 907

\bibitem{Jacobsen2007}
R. B. Jacobsen, C. A. Z. Vasconcellos, B. E. J. Bodmann, F. Fernandez,\textit{
Quark-gluon plasma in neutron stars}, Astron. Rel. Astrophys. (2010) 55

\bibitem{SuperMassiveNS}
Manuel Linares, \textit{Super-Massive Neutron Stars and Compact Binary Millisecond Pulsars}, PoS Proc. Sci., \textbf{362} (Nov 2020)

\bibitem{GTWC3_2021}
R. Abbott \textit{et. al.} [LIGO Scientific Collaboraton, Virgo Collaboration, KAGRA Collaboration],\textit{
GWTC-3: Compact Binary Coalescences Observed by LIGO and Virgo During the Second Part of the Third Observing Run}, arXiv:2111.03606
 
\bibitem{BinaryBHNS}
R. Abbott \textit{et. al.}, \textit{GW190814: Gravitational Waves from the Coalescence of a 23 Solar Mass Black Hole with a 2.6 Solar Mass Compact Object}, Astrophys. J. Lett., \textbf{896} (2020) L44

\bibitem{Romani2022}
R.W. Romani, D. Kandel, A.V. Filippenko, T.G. Brink, W. Zheng, \textit{PSR 0952-0607: The Fastest and Heaviest Known Galactic Neutron Star}, Astrophys. J. Lett., \textbf{934} (2022) L17

\bibitem{PhotoAstrometric2022}
C. Y. Lam \textit{et. al.}, \textit{An Isolated Mass-gap Black Hole or Neutron Star Detected with Astrometric Microlensing}, Astrophys. J. Lett, \textbf{933} (2022) L23

\bibitem{OGLE}
A. Udalski, A., M. K. Szyma\'{n}ski, and G. Szyma\'{n}ski, \textit{OGLE-IV: Fourth Phase of the Optical Gravitational Lensing Experiment}, Ac. Astro., \textbf{65 (1)} (2015) 1

\bibitem{MOA1}
J. B. Hearnshaw \textit{et. al.}, \textit{The MOA 1.8-metre alt-az Wide-field Survey Telescope and the MOA Project}, The 9th Asian-Pacific Regional IAU Meeting, held in Nusa Dua, Bali, Indonesia, Institut Teknologi Bandung Press (2006) 272


\bibitem{MOA2}
T. Sumi, \textit{MOA-II microlensing survey}, Proceedings of the Manchester Microlensing Conference: The 12th International Conference and ANGLES Microlensing Workshop (2008), eds. E. Kerins \textit{et. al.}

\bibitem{Sahu2022}
K. C. Sahu \textit{et. al}, \textit{An Isolated Stellar-mass Black Hole Detected through Astrometric Microlensing}, Astrophys. J., \textbf{933} (2022) 83

\bibitem{Mroz2022}
P. M\'{r}oz, A. Udalski, and A. Gould, \textit{Systematic Errors as a Source of Mass Discrepancy in Black Hole Microlensing Event OGLE-2011-BLG-0462}, arXiv:2207.10729v2

\bibitem{Thompson2019}
T. A. Thompson \textit{et. al.}, \textit{A noninteracting low-mass black hole-giant star binary system}, Science, \textbf{366} (2019) 637



\bibitem{Heuvel}
E. P. J. van den Heuvel and T. M. Tauris, Comment on ``A noninteracting low-mass black hole–giant star binary system", Science, \textbf{368} (2020) eaba3282


\bibitem{BH-NS1}
Y. Shao, \textit{On the Neutron Star/Black Hole Mass Gap and Black Hole Searches}, arXiv:2210.00425

\bibitem{BH-NS2}
N. Kumar, V. V. Sokolov, \textit{Mass Distribution and ``Mass Gap" of Compact Stellar Remnants in Binary Systems}, Astrophys. Bull., \textbf{77} (2022) 197  

\bibitem{DeSa2022}
L. M. de S\'{a}, \textit{Quantifying the evidence against a mass gap between black holes and neutron stars}, arXiv:2211.01447v1

\bibitem{Gao2023}
G. Shi-Jie, L. Xiang-Dong,\textit{ Can cosmologically-coupled mass growth of black holes solve the mass gap problem?}, arXiv:2307.10708v1

\end{thebibliography}
\end{document}